\documentclass[pdftex]{webofc}
\usepackage[varg]{txfonts}   
\RequirePackage{amsfonts}
\RequirePackage{amsmath,amssymb}
\RequirePackage{bm}
\usepackage{fancyhdr, graphicx}
\usepackage{float}
\usepackage{color}
\usepackage{amsmath,amssymb,wasysym}
\usepackage{epstopdf}
\usepackage{lineno}
\usepackage{url}
\usepackage[utf8]{inputenc}

\setbox123\hbox{\small$0$}
\def\S{\hbox to\wd123{\hss}}
\setbox124\hbox{\small$_{0}$}
\def\s{\hbox to\wd124{\hss}}

\newif\ifFirstAuthor
\FirstAuthortrue

\def\AddAuthor#1#2#3#4{%
        \def\PriAf{#2}%
        \def\SecAf{#3}%
        \def\ExtAf{#4}%
        \def\empty{}%
        \ifFirstAuthor
                \FirstAuthorfalse
        \else
                ,
        \fi
        \ifx\PriAf\empty
                #1\Aref{#4}%
        \else
                \ifx\SecAf\empty
                        \ifx\ExtAf\empty
                                #1\Iref{#2}%
                        \else
                        \ifx\ExtAf\empty
                                #1\IIref{#2}{#3}%
                        \else
                                #1\IIAref{#2}{#3}{#4}%
                                \relax
                        \fi
                \fi
        \fi
}

\def\AddInstitute#1#2{%
        \expandafter\write1{\string\newlabel{#1}{{#1}{}}}%
        \hbox to\hsize{\strut\hss$^{#1}$#2\hss}%
}

\begin{document}
\title{Recent Results from the CERN LHC Experiment TOTEM}
%
%
\subtitle{Implications for Odderon Exchange}

\author{\firstname{T.} \lastname{Cs\"{o}rg\H{o}, for the TOTEM Collaboration}\inst{1,2,3}\fnsep\thanks{\email{tcsorgo@cern.ch}} 
}
%
%

\institute{MTA Wigner FK, H-1525 Budapest 114, P.O.Box 49, Hungary
\and
          EKE KRC, H-3200 Gy\"ongy\"os, M\'atrai \'ut 35, Hungary 
          \and
          CERN, CH-1211 Geneva 23, Switzerland
          }
  



\abstract{%
Recent results are summarized from the TOTEM experiment at CERN LHC, including measurements of the total, elastic and inelastic cross-sections, the nuclear slope parameter $B$, the differential cross-section of elastic scattering and the 
real to imaginary part ration $\rho$ at $\sqrt{s} = $ 2.76  and 13 TeV. The implications of these data for Odderon (odd-gluon colorless) exchange are discussed.
}
\maketitle
\section{Introduction}
\label{s:introduction}
Currently, seven experiments are completing their data taking and data analysis programmes at the Large Hadron Collider (LHC), the most energetic particle accelerator made by humans so far. The four general purpose experiments, ALICE, ATLAS, CMS and  LHCb are supplemented by three specialized experiments, LHCf, MoEDAL and TOTEM, that focus on forward physics and search for exotic particle states. Five out of these seven experiments, namely ALICE, ATLAS, CMS, LHCb and TOTEM are overseen by the Resource Review Boards of  LHC. This report summarizes recent results from the TOTEM experiment (TOTal cross-section and Elastic scattering Measurement), presented at the XLVIII International Symposium on Multiparticle Dynamics, Singapore, in September 2018. In this work, we quote the final TOTEM results, thus this manuscript can also be considered as a brief summary of the results of the four most recent TOTEM publications of refs.~\cite{Antchev:2017dia,Antchev:2017yns,Antchev:2018edk,Antchev:2018rec}.

\section{TOTEM physics and experimental setup}
\label{s:setup}

In general, the goal of the TOTEM experiment is to measure colorless exchange, including elastic, single and double diffractive scattering as well as central exclusive production at the energies of the Large Hadron Collider at CERN. These processes correspond to an increasing elastic  fraction of the total cross-section and their precise measurement requires special experimental setup in the forward direction, in the LHC tunnel extending as far as 220 m from Interaction Point 5 (IP5), the collision point that is used for measurements by both the CMS and the TOTEM experiments. By the time of writing this manuscript, a common CMS-TOTEM precision proton spectrometer or CT-PPS project, that started as a common CMS - TOTEM effort, has been fully integrated to the CMS experiment and became the PPS project of CMS. Although this report summarizes the recent standalone TOTEM results, it is important to mention that the (CT-)PPS project allowed to operate the world's most complex Roman Pot detector system under regular LHC running conditions and resulted in a successful CMS-TOTEM data taking period during LHC Run-2 (2016-2018) with data recorded with tagged forward protons exceeding 100 fb$^{-1}$. The harvesting of the rich physics potential of the (CT-)PPS dataset has just been started~\cite{Cms:2018het}.

The TOTEM experimental setup consists of two inelastic telescopes T1 and T2 to detect charged particles coming from inelastic $\rm pp$ collisions and the Roman Pot detectors (RP) to detect elastically scattered protons at very small angles~\cite{Anelli:2008zza}.
A RP unit consists of 3 RPs, two approaching the outgoing beam vertically and one horizontally.
Each RP is equipped with a stack of 10 silicon strip detectors designed with the specific objective of reducing the insensitive area at the edge facing the beam to only a few tens of micrometers. 
The $5.4$~m ($7$ m) long lever arm between the near and the far RP units at $\sqrt{s} = 2.76 $ TeV (13 TeV) has the important advantage that the local track angles in the $x$ and $y$-projections perpendicular to the beam direction can be reconstructed with a precision of 2~$\mu$rad (3 ~$\mu$rad), respectively.

The inelastic telescopes are placed symmetrically on both sides of Interaction Point 5 (IP5): the T1
telescope is based on cathode strip chambers (CSCs) placed at $\pm$9~m and covers the pseudorapidity range 3.1~$\le |\eta| \le$~4.7; the T2 telescope is based on gas electron multiplier (GEM) chambers placed at $\pm$13.5~m and covers the pseudorapidity range 5.3~$\le |\eta| \le$~6.5.  The pseudorapidity coverage of the two telescopes at $\sqrt{s}=2.76$~TeV allows the detection of about 92~\% of the inelastic events.  As the fraction of events with all final state particles beyond the instrumented region has to be estimated using phenomenological models, the
excellent acceptance in TOTEM minimizes the dependence on such models and thus provides small uncertainty on the inelastic rate measurement.

The complete and detailed description of the TOTEM experiment is given in refs~\cite{Anelli:2008zza,TOTEM:2013iga}. 
Specific details of the experimental setups used for the recent TOTEM measurements at $\sqrt{s}= 2.76$ and $13$ TeV of the total, elastic and inelastic cross-sections, the nuclear slope parameters
are given in refs.~\cite{Antchev:2017dia,Antchev:2017yns,Antchev:2018edk,Antchev:2018rec}. 
Below,  we detail only some of the specific modifications of the general TOTEM experimental setup, that were necessary to for the data taking and to achieve the physics  goals of the measurements at $\sqrt{s} = 2.76 $ and $13 $ TeV, respectively.

\section{Data taking}
\label{s:data-taking}

The Roman Pot (RP) units used for the measurement at  $\sqrt{s} = 2.76 $ TeV are located on both sides of the IP at distances of $\pm214.6$~m (near) and $\pm220.0$~m (far) from IP5~\cite{Antchev:2018rec}.
The analysis is performed on a data sample (DS1-2.76) recorded in 2013 during an LHC fill with
$\beta^* = 11$ m injection optics.  The RP detectors were inserted to 13 times the transverse beam size. Although the differential cross-section measurement is based on the analysis of this DS1-2.76 dataset, 
the  total cross-section measurement at $\sqrt{s} = 2.76$ TeV is based on a different data set (DS2-2.76) that was recorded with at similar beam conditions but RP detectors placed at 4.3 times the transverse
beam size.  This data set DS2-2.76 had higher statistics and it was also used in order to obtain the  final normalization of the data set DS1-2.76, as detailed in ~\cite{Antchev:2018rec}.
The horizontal RP detectors were not inserted during the data taking at $\sqrt{s} = 2.76$ TeV, and the vertical alignment uses the RP position sensors and is further refined with precise constraints based on symmetries of elastic scattering~\cite{Nemes:2017gut}.

The RP units used for the measurements at $\sqrt{s} = 13$ TeV are located on both sides of the LHC Interaction Point 5 (IP5) at distances of $\pm213$~m (near) and $\pm220$~m (far) ~\cite{Antchev:2017dia}.
At $\sqrt{s} = 13$ TeV, the  horizontal RP detectors were inserted and their  overlaps with the two vertical RPs allowed for a precise relative alignment of the detectors within the unit. 

At $\sqrt{s}= 13$ TeV, the data analysis has been performed on a large data sample, including seven data sets (DS1 - DS7) recorded in 2015 during a special LHC fill with $\beta^* = 90$ m optics and detailed in ref.~\cite{Antchev:2017dia}.  The RP detectors were placed at a distance of 10 times the transverse beam size ($\sigma_{\mbox{\rm beam}}$) from the outgoing
beams. The special trigger settings allowed to collect about $10^9$
elastic events. The angular resolution was different for each of the data sets DS1-DS7, and it deteriorated with time within the fills, expected mainly due to the beam emittance growth.
The data sets have been reorganized according to their resolution into two larger data sets. The ones with better (about 20 \%) resolution were collected into DS$_g$, which includes DS1, DS2 and DS4. The remaining ones are collected in data set DS$_o$. 
The normalization of the differential cross-section measurement at $\sqrt{s} = 13$ TeV
is based on the total cross-section measurement at the same energy. The total cross-section 
analysis was performed on a data set DS$_n$,  that was also measured with a $\beta^* = 90$ m optics, 
but with the RP detectors placed two times closer to the beam, to 5 times the transverse beam size.

We also report on the measurement of the ratio of the real to imaginary part of the forward scattering amplitude, the parameter $\rho$ obtained from a special measurement in the Coulomb-Nuclear Interference (CNI) region~\cite{Antchev:2017yns}. 
This measurement was based on data taken in September 2016 during a sequence of dedicated
LHC proton fills with the special beam properties corresponding to $\beta^* = $ 2500 m.
The vertical RPs approached the beam centre to only about 3 times the vertical beam width, $\sigma_y$, corresponding roughly to 0.4 mm. 
Such an  exceptionally close distance was required in order to reach very low $|t|$ values
and was possible due to the low beam intensity in this special beam operation: each beam contained only
four or five colliding bunches and one non-colliding bunch,  with about 5 $\times 10^{10}$ protons in each bunch~\cite{Antchev:2017yns}.
The horizontal RP-s were  used for the track-based alignment only, and therefore
they were placed at a safe distance of 8 times the horizontal beam with, corresponding to about 5 mm.
This horizontal distance was close enough to have the horizontal RP overlapping with the vertical RPs
~\cite{Antchev:2017yns}.
Let us also stress that TOTEM measurements provided a set of consistent values for the total cross-section of proton-proton scattering using three different methods (the inelastic-independent method of refs. ~\cite{Antchev:2011vs} and ~\cite{Antchev:2013gaa}, 
 the $\rho$-independent method of ref.~\cite{Antchev:2013haa} and the luminosity-independent method of ref.~\cite{Antchev:2013iaa} 
 at 7 TeV, yielding values of $\sigma_{\mbox{\rm tot}} = 98.3 \pm 2.0$ mb, $98.6  \pm 2.3$ mb, as well as $99.1 \pm 4.3$ mb and $98.1 \pm 2.4 $ mb, respectively. Similar measurements at 8 TeV indicated that our results for the total proton-proton cross-section
are stable not only for the choice of the method,   also  for very different beam conditions as well~\cite{Antchev:2013paa}.
The results for various TOTEM cross-section measurements, including the recent results at $\sqrt{s}= 2.76$ and $13$ TeV are summarized in Table~\ref{f:stot-el-inel-s}, and are discussed in greater details in subsection~\ref{ss:cross-sections}, based on  a recent and more detailed TOTEM review~\cite{Antchev:2017dia}.

\section{Elastic analysis}
\label{s:analysis}
The horizontal and vertical scattering angles of the proton at IP5 ($\theta^*_x$,$\theta^*_y$) are reconstructed in
a given arm by inverting the proton transport equations using TOTEM's special LHC optics reconstruction and recalibration method, detailed in ref.
~\cite{Antchev:2014voa}. The scattering angles obtained for the two arms are
averaged and the four-momentum transfer squared is calculated as $t = - p^2 \theta^{*2}$, where
$p$ is the LHC beam momentum and the scattering angle $\theta^{*} = \sqrt{\theta_x^{*2} + \theta_y^{*2} }$.
Precise understanding of the proton transport at IP5 is of key importance for the success of the TOTEM experiment. 
In all the TOTEM analysis presented here, a novel method of optics evaluation is utilized,
based on ref.~\cite{Antchev:2014voa}, which exploits the kinematic constraints of  elastically scattered protons 
observed in the RPs. Typically we find that the residual uncertainty of the optics estimation method is smaller than $0.25 \%$,
which makes it possible to determine the total cross-sections with about 2-3 \% relative precision.

\begin{figure*}
\centering
\includegraphics[width=0.48\textwidth]{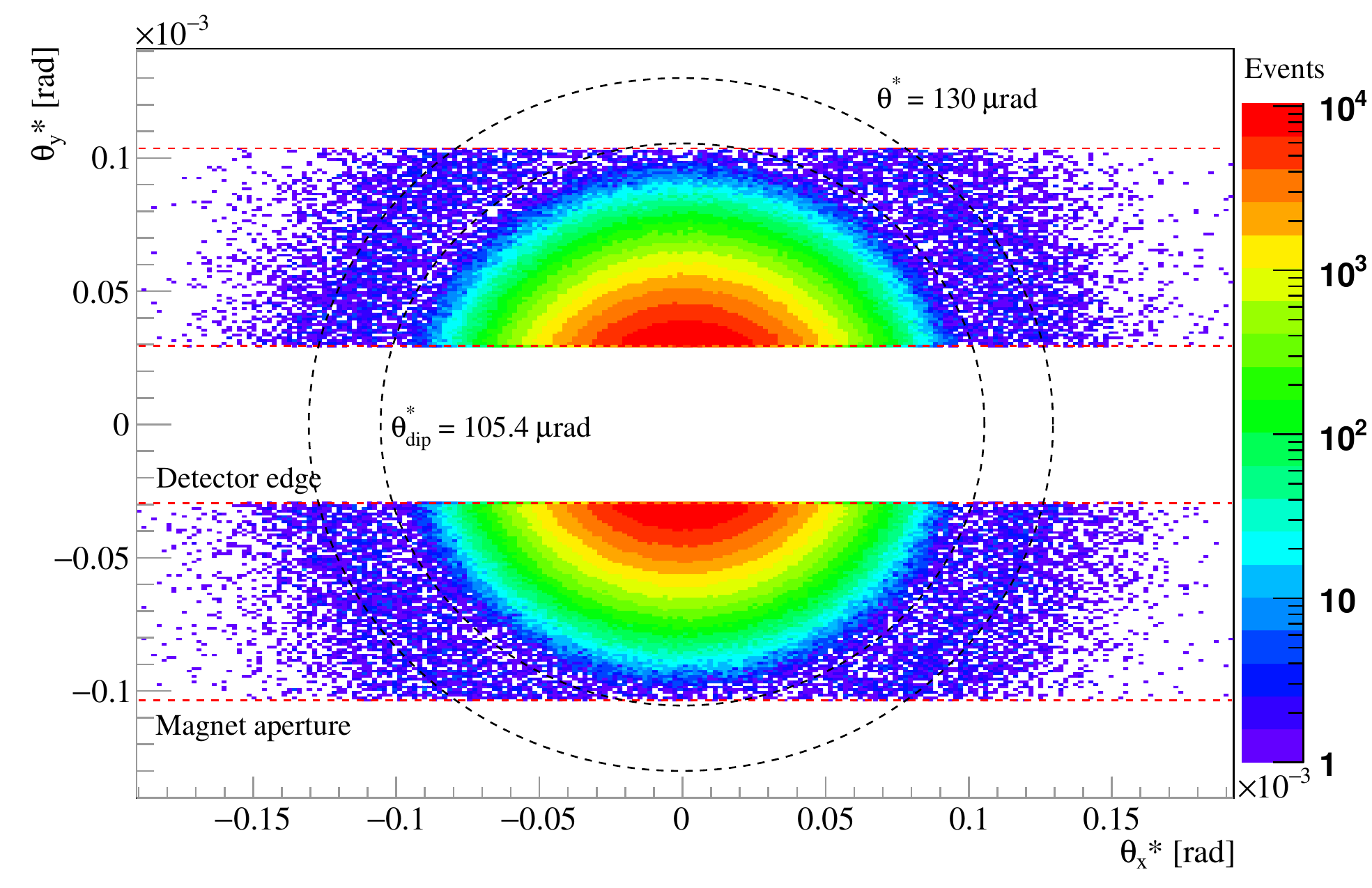}
\includegraphics[width=0.48\textwidth]{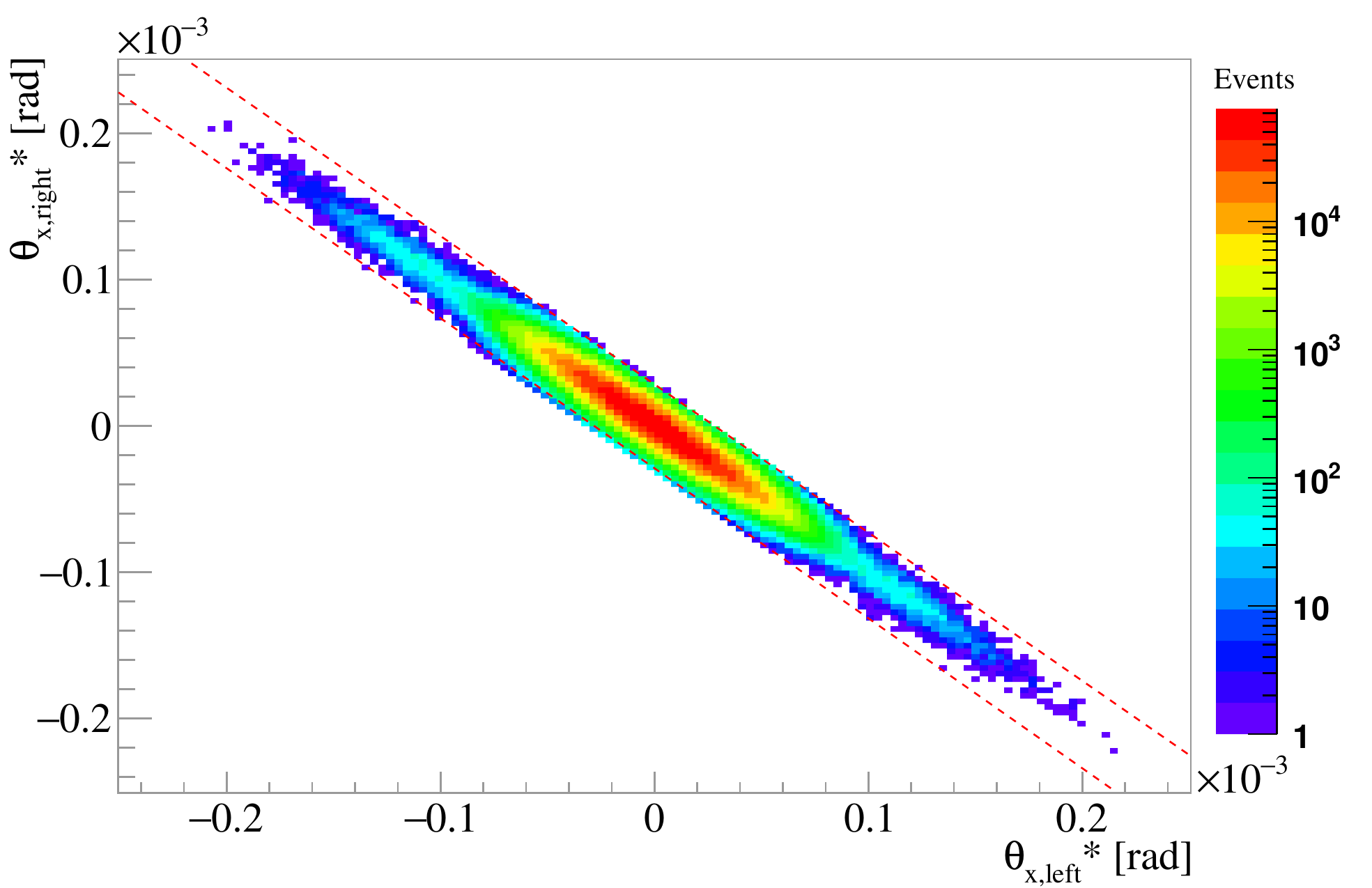}
\caption{
{\bf Left panel:}
Rotational symmetry of elastic scattering in the $(\theta_x^*,\theta_y^*)$ plane at $\sqrt{s} = 13$ TeV proton-proton collisions, 
as measured by TOTEM and detailed in ref.~\cite{Antchev:2018edk}.
The red dashed lines show the analysis acceptance cuts,
which define the acceptance boundaries near the detector edge and magnet aperture. 
The inner black dashed circle illustrates the approximate scattering angle position
$\theta_{dip}^* = 105.4$ $\mu \mbox{\rm rad}$  of the diffractive minimum, 
while the outer black dashed line indicates the approximate scattering angle position of
$\theta_{bump}^* = 130$ $\mu \mbox{\rm rad}$  of the diffractive maximum. 
{\bf Right panel:}
Collinearity of the horizontal scattering angles at the collision point $\theta_{x, \mbox{\rm  \scriptsize left}}^*$ 
and $\theta_{x, \mbox{\rm \scriptsize right}}^*$ in elastic $pp $ scattering at $\sqrt{s} = 13$ TeV proton-proton collisions, as reconstructed by TOTEM~\cite{Antchev:2018edk}. The dotted lines indicates
4$\sigma$ physics cuts applied to select the elastic events. See ref.~\cite{Antchev:2018edk} for
the collinearity cuts on the vertical scattering angle $\theta^*_y$ as well as further details on the other analysis cuts.
}
\label{f:fig1}       
\end{figure*}


Instead of detailing this method, let me here just highlight some of its beautiful applications and results.
The azimuthally uniform distribution of the scattering angle $\theta^{*}$ demonstrates the azimuthal symmetry of
elastic scattering. This is illustrated on Figure~\ref{f:fig1}, which shows the analysis acceptance cuts
with red dashed lines, near to the edge of the detector and the aperture of the LHC magnets. On Figure~\ref{f:fig1} two black circles highlight the quantum interference  apparent in these data, that reveal a local scattering minimum and maximum behaviour. 
These rings, visible by eye even on the raw TOTEM data, clearly indicate a quantum interference pattern in  elastic proton-proton scattering at LHC energies. These rings suggests that protons have a composite structure that can be directly observed by quantum interference in elastic proton-proton scattering at LHC energies, see ref.~\cite{Antchev:2018edk} for further details on the corresponding differential elastic cross-section data.

The right panel of Figure~\ref{f:fig1} indicates the collinearity of the horizontal scattering angle $\theta_x^*$
in the left and right going directions in elastic $pp $ scattering at $\sqrt{s} = 13$ TeV proton-proton collisions, as reconstructed by TOTEM
~\cite{Antchev:2018edk}. The dotted lines indicates
4$\sigma$ physics cuts applied to select the elastic events. Ref.~\cite{Antchev:2018edk} details
the collinearity cuts on the vertical scattering angle $\theta_y^*$ as well as the other analysis cuts that were defined to select progressively the elastic events.

The selection of elastic events and the main steps of the analysis were similar at 13 and 2.76 TeV:
both data analysis started with the reconstruction of kinematics, detector alignment, recalibration of the LHC optical functions with the constraints coming from the measured pp  elastic scattering data. The measured differential cross-sections were corrected for resolution unfolding and acceptance, background substraction and detection efficiency, angular resolution, normalization and binning effects ~\cite{Antchev:2018rec,Antchev:2018edk}.
The horizontal and relative near-far alignment was done based on the observed tracks. The analysis at 2.76 TeV was further complicated by the lack of the horizontal RP-s, that made track-based bottom-top RP alignment impossible, so new methods were developed for absolute $y$-alignment of the two diagonals.  These were based on two constraints from the symmetry of elastic scattering. The first constraint was that the barycenter of the distribution of the $\theta^*_y$ scattering angle was aligned to zero. The second constraint was that after rescaling the distribution of the $\theta^*_x$ and $\theta^*_y$
horizontal and vertical scattering angles should be the same ~\cite{Antchev:2018rec}. Fortunately the calibration of the LHC optics was independent from the detector alignment procedure.

\section{Results at $\sqrt{s} = 13$ and $2.76$ TeV}

\subsection{Measurement of $(\rho,\sigma_{\rm tot})$ at 13 TeV -- implications for Odderon exchange \label{ss:rho}}
\label{ss:results-rho}
The TOTEM experiment at the LHC has measured the differential elastic proton-proton scattering cross section down to $|t| = 8 \times 10^{-4}$ GeV$^2$ at the centre-of-mass energy of $\sqrt{s} = 13$  TeV, using a special LHC optics with  
$\beta^* = 2.5$ km, as detailed in ref.~\cite{Antchev:2017yns}. 
This allowed TOTEM to access the Coulomb-nuclear interference (CNI) region and to determine $\rho$, 
the real-to-imaginary ratio of the hadronic scattering  amplitude at $t=0$ with an unprecedented precision. 


Measurements of the total proton-proton cross-section and $\rho$ have been published in the literature from the low energy range of $\sqrt{s} \approx 10$ GeV up to the LHC energy of 8 TeV~\cite{Patrignani:2016xqp}. 
Such experimental measurements have been parametrised by a large variety of phenomenological models in the last decades, and
were analysed and classified by the COMPETE collaboration ~\cite{Cudell:2002xe}.

\begin{figure*}
\centering
\includegraphics[width=0.90\textwidth]{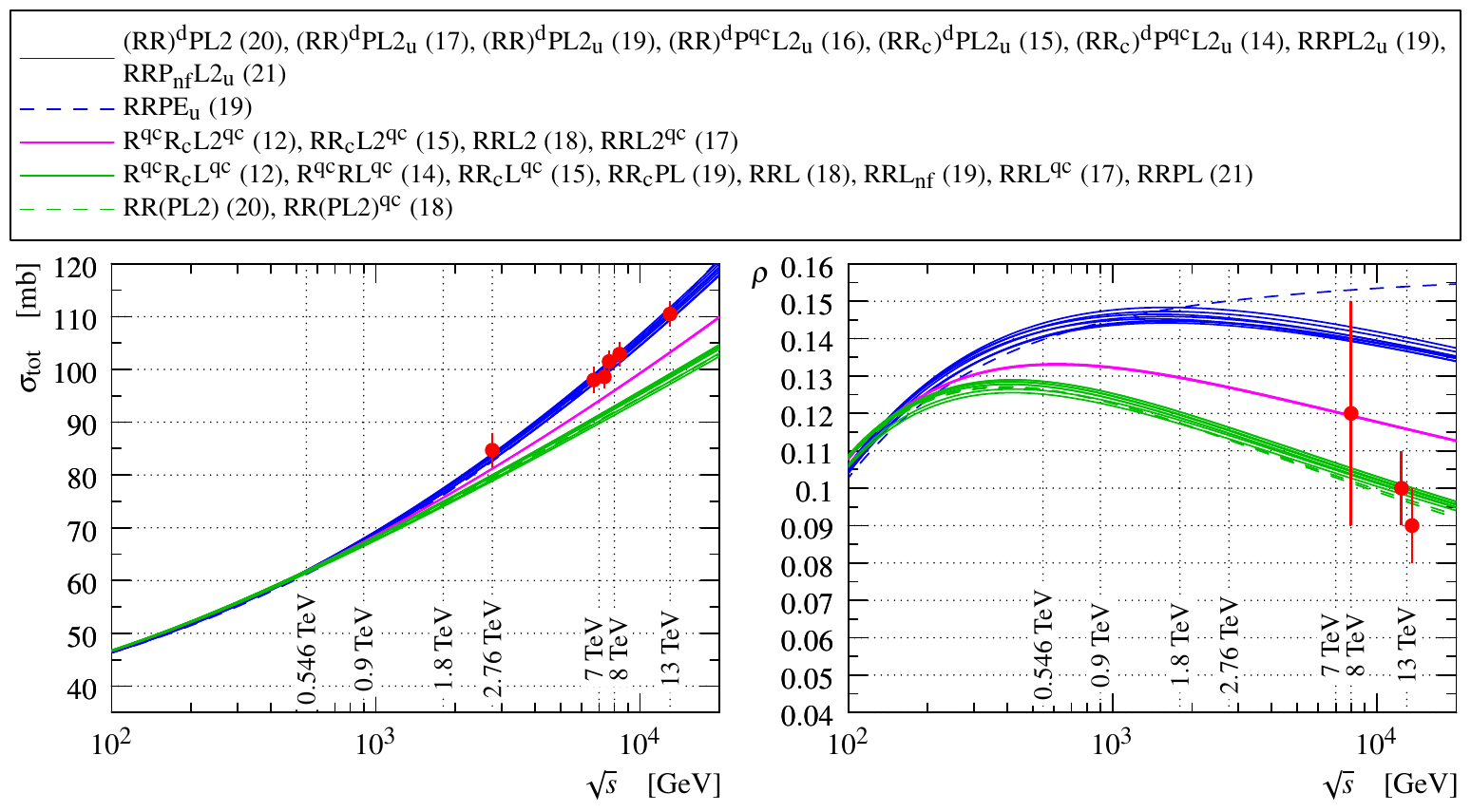}
\caption{
COMPETE bands for the total cross-section $\sigma_{tot }(s)$
and the $\rho(s)$ indicate~\cite{Cudell:2002xe} that without an Odderon contribution, 
recent TOTEM data at $\sqrt{s} = 13$ TeV for the pair of
$(\sigma_{tot },  \rho)$
cannot  be described simultaneously. 
}
\label{f:rho-sigmatot-s}       
\end{figure*}

One of the most inspiring recent observation of TOTEM indicates the 
presence of a crossing-odd component in the scattering amplitude of $pp$ and $p\overline{p}$ elastic collisions at the LHC energies, the so called Odderon effect, proposed in 1973 by Lukaszuk and Nicolescu~\cite{Lukaszuk:1973nt}.
Figure ~\ref{f:rho-sigmatot-s} of TOTEM~\cite{Antchev:2017yns} indicates one of the indirect Odderon effects. 
This Figure  clearly demonstrates,  that none of the models considered by COMPETE are able to describe simultaneously,
without taking into account a crossing-odd component of the scattering amplitude,
the TOTEM $\rho$  measurement at $\sqrt{s} = 8$ and $13$ TeV together with the ensemble of the total cross-section measurements by TOTEM in the $\sqrt{s} =   2.76$ to $13$ TeV energy range~\cite{Antchev:2013iaa,Antchev:2015zza,Antchev:2016vpy,Antchev:2017dia}.
The exclusion of the COMPETE published models is clearly illustrated by  Figure~\ref{f:rho-sigmatot-s}.
The same quantitative and qualitative conclusion is reached  with a  $p$-value analysis
as detailed in ref.~\cite{Antchev:2017yns}. 

\begin{figure*}
\centering
\includegraphics[width=0.90\textwidth]{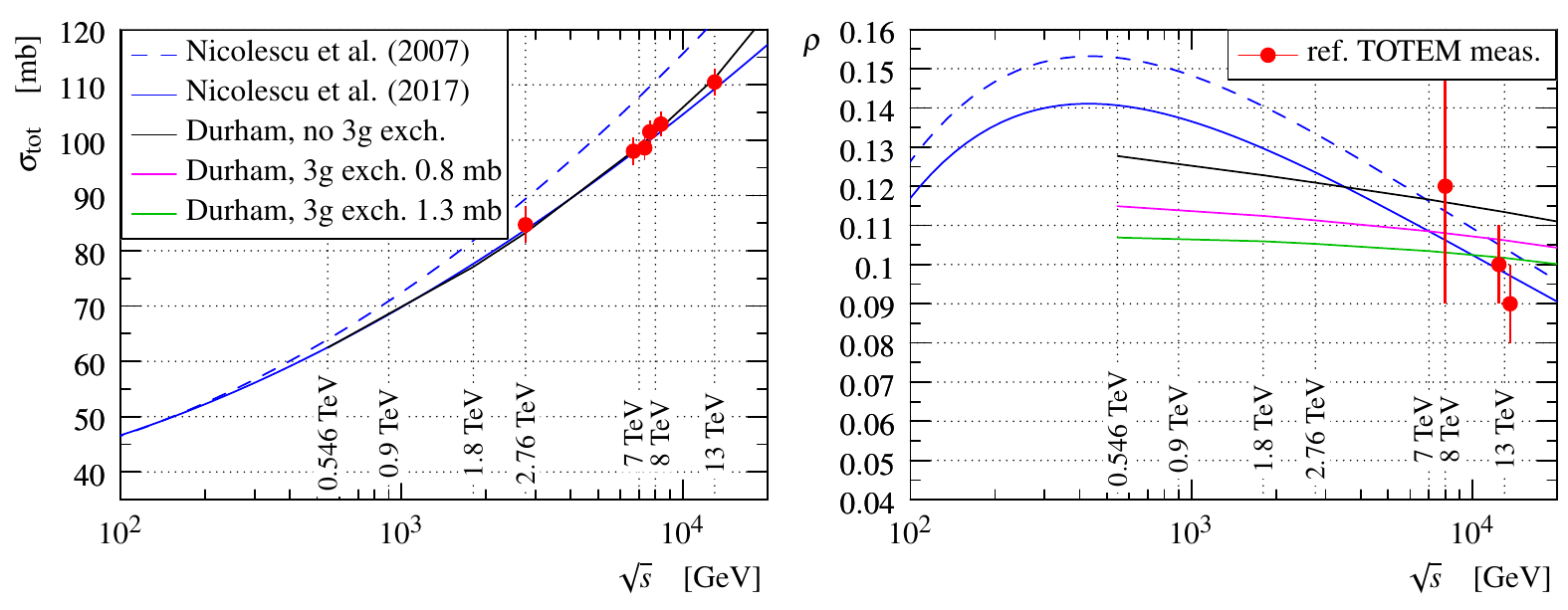}
\caption{
Measured values of  $\sigma_{tot}$ and $\rho$ are compared to two different class of model calculations. 
The comparisons indicate that with an Odderon contribution, the recent TOTEM data  for the pair of 
$(\sigma_{\rm tot},\rho)$ at $\sqrt{s} = 13$ TeV can be described simultaneously:
Predictions of the Odderon model by Nicolescu and collaborators are shown by a dashed~\cite{Avila:2006wy} and solid blue curves~\cite{Martynov:2017zjz}.
Results from the KMR model~\cite{Khoze:2017swe} without Odderon effects  are shown with a black line, and with Odderon effects are shown with magenta and green lines, 
where  the crossing-odd contribution is taken from ref.~\cite{Levin:1990gg}. 
Reference TOTEM measurements are indicated by red dots. 
}
\label{f:rho-sigmatot-comparisons}       
\end{figure*}

The presence of an Odderon effect in the pair of excitation functions $(\rho(s),\sigma_{\rm tot}(s)) $ 
is further supported by Figure ~\ref{f:rho-sigmatot-comparisons},
that compares the measured values of $\rho$ and $\sigma_{tot}$ to two different class of model calculations, indicating the Odderon exchange effects explicitely and directly. 
Predictions of a model by Nicolescu and collaborators
from refs.~\cite{Avila:2006wy,Martynov:2017zjz} together with the
the Durham or KMR model of Khoze, Martin and Ryshkin 
~\cite{Khoze:2017swe} (that also
included a crossing-odd contribution from ref.~\cite{Levin:1990gg})
were compared to the reference TOTEM
measurements (red dots). 
These results, summarized on Figure~\ref{f:rho-sigmatot-comparisons}, confirm that
at $\sqrt{s} = 13 $ TeV, the pair of $(\rho, \sigma_{\rm tot})$ data is 
best described with the help of Odderon effects~\cite{Antchev:2017yns}.

\subsection{Total, elastic and inelastic cross-sections}
\label{ss:cross-sections}
Fig.~\ref{f:dsdt-2.76} indicates TOTEM results for the differential cross-section of elastic pp scattering at $\sqrt{s} = 2.76 $ TeV.
The low-$t$ part of the measured distribution is frequently approximated with an exponential,
\begin{equation}
    \frac{d\sigma}{dt} = A \, \exp(B t), \label{e:dsdt-exp}
\end{equation}
with the values of $B$ and $R$ indicated on Fig.~\ref{f:dsdt-2.76}.
The $B$ and $R$ results of TOTEM  are  summarized for $\sqrt{s} = 2.76 $ TeV and $13$ TeV
in Table~\ref{t:BR}. Here ratio R is defined as $ R= max/min$, 
this quantity characterizes the dip-bump region of the differential cross-section of elastic scattering
beyond the domain of eq.~(\ref{e:dsdt-exp}).
The ratio $R$ is the ratio of the  value of the differential cross section at the (first) diffractive maximum and minimum, 
denoted here as $max$ and $min$, respectively.

Two comments are due. The first of these comments is that eq.~(\ref{e:dsdt-exp}) corresponds to  an exponential ``diffractive cone" approximation, that may be valid in the low-$t$ domain only. This equation corresponds to the so called ``Grey Gaussian" approximation, that suggests a relationship between the nuclear slope parameter $B$, the real to imaginary ratio $\rho_0$, the total cross-section $\sigma_{\rm tot}$ and the elastic cross-section
$\sigma_{\rm el}$ as follows~\cite{Block:2006hy,Fagundes:2011hv,Broniowski:2018xbg}:
\begin{equation}
    A \, = \, B \, \sigma_{\rm el} \, = \, \frac{1+\rho_0^2}{16 \, \pi}\, \sigma_{\rm tot}^2, \qquad\qquad
    B \, = \, \frac{1+\rho_0^2}{16 \, \pi }\, \frac{\sigma_{\rm tot}^2}{\sigma_{\rm el}}.~\label{e:ABsigma}
\end{equation}
Let us note that in the present notation we suppress the $s$-dependence of the
observables, ie. $\sigma_{\rm tot} \equiv \sigma_{\rm tot}(s)$, $\sigma_{\rm el} \equiv \sigma_{\rm el}(s)$ etc.
The above relationships, in a slightly modified form, have been utilized by TOTEM to measure the total cross-section
by TOTEM using the luminosity independent method at 2.76, 7, 8 and 13 TeV in refs.~\cite{Nemes:2017gut,Antchev:2013iaa,Antchev:2013paa,Antchev:2017dia}, respectively, based on the following luminosity independent formula:
\begin{equation}
    \sigma_{\rm tot}  =  \sigma_{\rm el} + \sigma_{\rm inel} \, = \, \frac{16 \, \pi}{1+\rho_0^2}\, \frac{\left. \frac{d\sigma}{dt}\right|_{t=0}}{\sigma_{\rm el}+\sigma_{\rm inel}}
       \, = \, \frac{16 \, \pi}{1+\rho_0^2}\, \frac{\left.\frac{dN}{dt}\right|_{t=0}}{N_{\rm el}+N_{\rm inel}}. \label{e:sigmatot-lumi}
\end{equation}
Expressing parameters $A$ and $B$ in terms of the elastic and the total cross-section as given in eq.~(\ref{e:ABsigma}) is particularly
useful, when we discuss  the inelastic profile function called also the shadow profile of the protons, as detailed below.

The second comment relates to the ratio of the elastic to the total cross-section, $\frac{\sigma_{\rm el}}{\sigma_{\rm tot}}$.
The shadow profile function is introcuded as $P(b) = 1 -|\exp\left[-\Omega(b)\right]|^2$, where $\Omega(b)$ is the 
so-called opacity function, which is generally complex. It is defined with the help of the
relation $t_{\rm el}(b) = i \Big( 1 - \exp\big[ -\Omega(b) \big] \Big) $, 
where $t_{\rm el}(b)$ stands for the Fourier-Bessel transformed elastic scattering amplitude $T_{\rm el}(\Delta)$, where $\Delta = \sqrt{-t}$ is the modulus of the four-momentum transfer in elastic scattering. For more details on these transformations and convention, see refs.~\cite{Bialas:2006qf,Nemes:2015iia,Csorgo:2018uyp}.
For clarity, let us note that other conventions are also used in the literature and for example the shadow profile  $P(b)$ is also referred to as the inelastic profile function as it corresponds to the probability distribution of inelastic proton-proton collisions in the impact parameter $b$ with $0\le P(b) \le 1$. When the real part of the scattering amplitude is neglected, $P(b)$ is frequently denoted as $G_{\rm inel}(s,b)$, see for example refs.~\cite{Petrov:2018wlv,Dremin:2013qua,Dremin:2014spa,Dremin:2018urc,Dremin:2019tgm}.

\begin{figure*}
\centering
\includegraphics[width=0.475\textwidth]{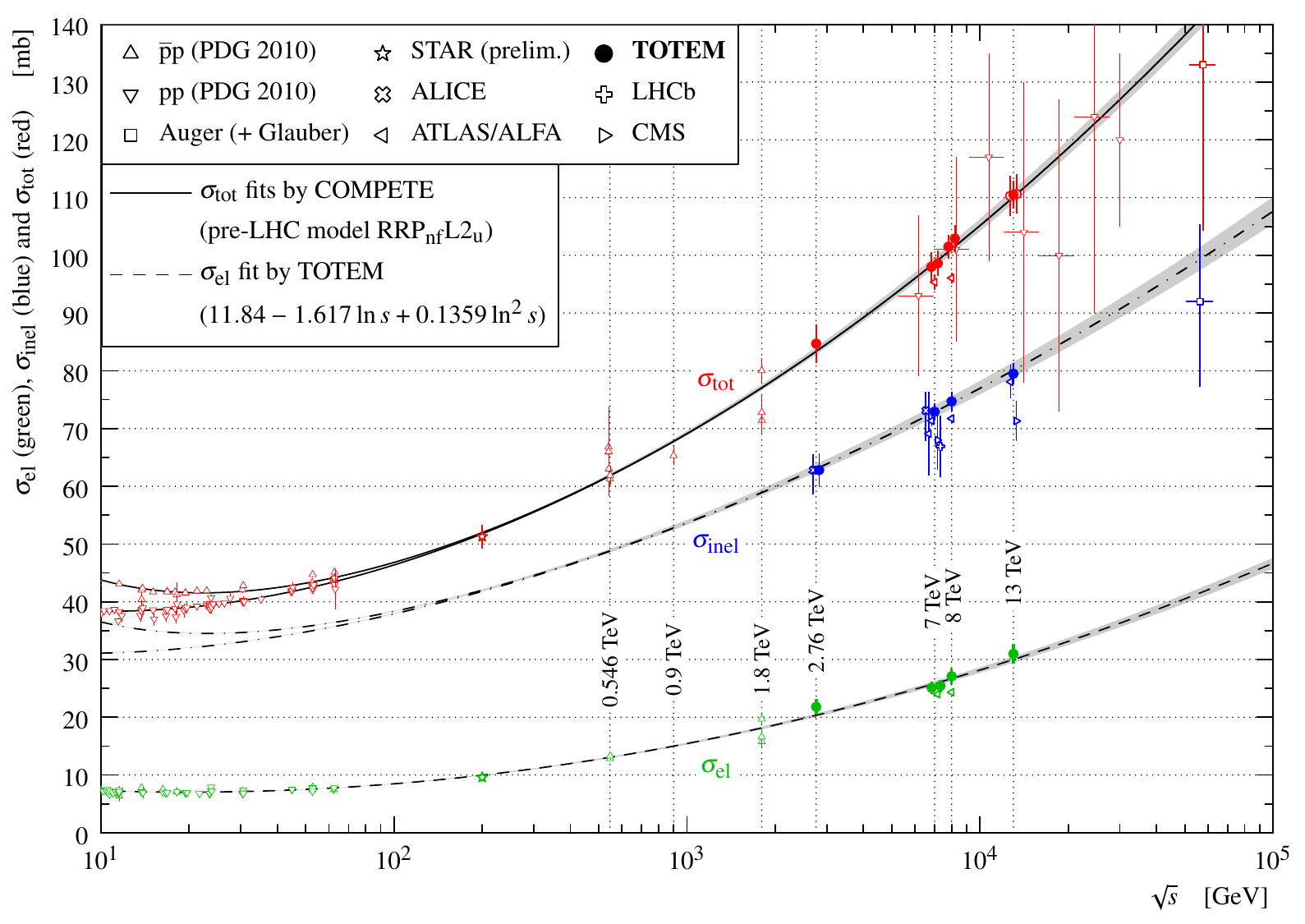}
\raisebox{-1ex}{\includegraphics[width=0.50\textwidth]{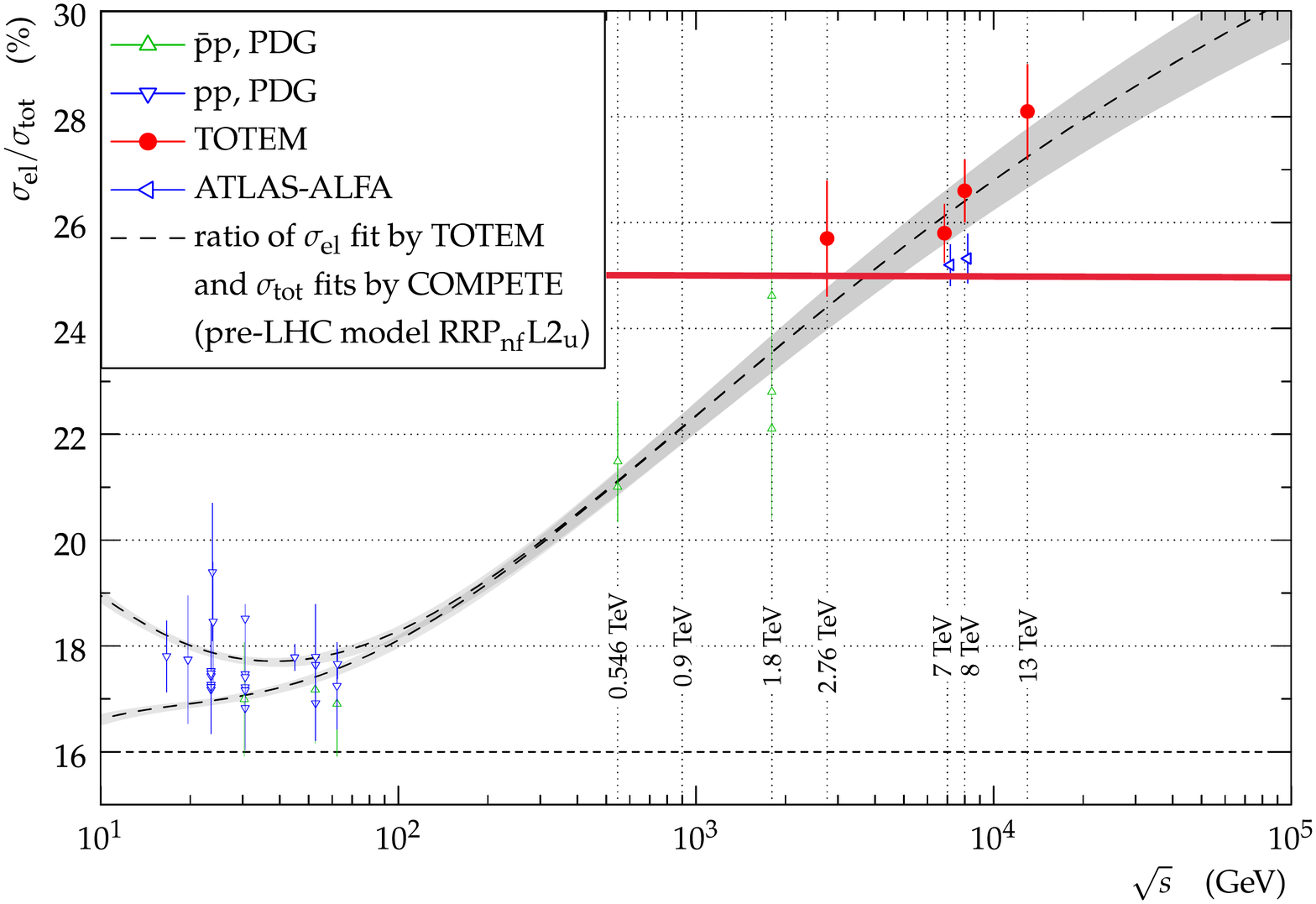}}
\caption{
{\bf Left panel:}
The total, elastic and inelastic cross-sections as measured at various LHC energies and below.
The total cross-section at $\sqrt{s} = 13 $ TeV is determined with $\rho = 0.1$ at this energy.
Overview of total ($\sigma_{tot}$),
inelastic ($\sigma_{el}$)
and elastic ($\sigma_{el}$) cross-sections 
for $pp$
and $p\overline{p}$  collisions as a function of 
$\sqrt{s}$, 
including TOTEM measurements over the whole energy
range explored by the LHC.
Uncertainty band on theoretical models
and/or fits are described in the Legend. 
The continuous black lines (lower for $pp$, upper for
$p\overline{p}$) represent the best fits of the total cross section data by the COMPETE collaboration.
The dashed line results from a fit of the elastic scattering data. The dash-dotted lines refer to
the inelastic cross section and are obtained as the difference between the continuous and dashed
fits. From refs. ~\cite{Antchev:2017dia,Antchev:2017yns}.
{\bf Right panel:}
The elastic to total cross-section ratio increases with increasing energies, as indicated on this Figure from
ref.~\cite{Antchev:2017dia}. At the LHC energies between 8 and 13 TeV it 
crosses significantly the important limit of $(1+\rho_0^2) \sigma_{el} / \sigma_{tot} =  1/4$. 
It is important to note that this ratio reaches the critical value in the region of $\sqrt{s} = 2.76 - 7$ TeV and
it clearly exceeds it at $\sqrt{s} = 13$ TeV. 
The best COMPETE fit suggests a threshold behaviour in the region of  $\sqrt{s} = 2.76$ - $4$ TeV.
}
\label{f:stot-el-inel-s}       
\end{figure*}

In the  exponential elastic cone approximation of eqs.~(\ref{e:dsdt-exp},\ref{e:ABsigma}), 
the shadow profile function has a remarkable and very interesting behaviour, as anticipated in ref.~\cite{Broniowski:2018xbg}:
\begin{equation}
    P(b)  =  1 - \Big[ 1 - r \, \exp\Big( - \frac{b^2}{2 B}\Big)\Big]^2 \, - \, \rho_0^2  r^2 \, \exp\Big( - \frac{b^2}{ B}\Big) , 
    \qquad\mbox{\rm where}\qquad r   \, = \, 4\, \frac{ \sigma_{\rm el}}{\sigma_{\rm tot}} . \label{e:Prb} 
\end{equation}
Thus the shadow profile at the center, $P_0 = P(b=0)$ reads as
\begin{equation}
P_0 \, = \,  \frac{1}{1+\rho_0^2} \, - \, (1+\rho_0^2) \, \Big[ r - \frac{1}{1+\rho_0^2}\Big]^2 ,
\end{equation}
which cannot become maximally absorptive or black ($P_0 = 1$)  at those colliding energies, where 
$\rho_0$ is not negligibly small. The maximal absorption corresponds to $P_0 \, = \,  \frac{1}{1+\rho_0^2}$, reached 
where the ratio of the elastic to total cross-sections reaches the value $r = 1/(1+\rho_0^2)$, corresponding to $ 4 \Big(1 + \rho_0^2\Big) \,\sigma_{\rm el} = \sigma_{\rm tot}$. 
Given that $\rho_0 \le 0.15$ as indicated on Figure~\ref{f:rho-sigmatot-s} and $\rho(s)$ seems to  decrease with increasing energies at least in  the 8 $\le \sqrt{s} \le 13$ TeV region, the critical value of the elastic to total cross-section ratio corresponds to about $\sigma_{\rm el}/\sigma_{\rm tot} \approx 24.5-25.0 $ \%. Table~\ref{t:2.76} and the right panel of Figure~\ref{f:stot-el-inel-s} indicates that this threshold, within errors, is reached approximately already at $\sqrt{s} = 2.76 $ TeV. The threshold behavior saturates somewhere between 2.76 and 7 TeV. According to the best COMPETE extrapolation,
as indicated on the right panel of Figure~\ref{f:stot-el-inel-s} , such a  transition may happen
around the threshold energy of $\sqrt{s_{\rm th}} \approx 2.76 - 4 $ TeV. As indicated on this right panel of  Figure ~\ref{f:stot-el-inel-s}, the elastic to total cross-section ratio becomes significantly larger than the threshold value at $\sqrt{s} = 13 $ TeV colliding energies.

It follows that the inelastic or shadow profile function of the proton undergoes a qualitative change
in the region of  $2.76 < \sqrt{s} < 7 $ TeV energies. 
The investigation of such a dip or hollowness, corresponding to 
$\sigma_{\rm el} \ge \sigma_{\rm tot}/4/ \Big(1 + \rho_0^2\Big)$ according to Equation~\ref{e:Prb}, is
a hotly debated, current topic in the literature.  At high energies, with $\sigma_{\rm el} \ge \sigma_{\rm tot}/4$,
hollowness may become a generic property of
the shadow profile functionsthat characterize the impact parameter distribution of inelastic scatterings. 
The maximum of $P(s,b=0) \approx 1$ at $\sigma_{\rm el}(s) \approx \sigma_{\rm tot}(s)/4$ 
seems to be rather independent of the detailed $b$-dependent  shape of the inelastic collisions, see for example ref.~\cite{Broniowski:2018xbg}. 
We recommend  refs.~\cite{Troshin:2007fq,Fagundes:2011hv,Dremin:2013qua,Alkin:2014rfa,Troshin:2014rva,Dremin:2014spa,Anisovich:2014wha}
for early papers as well as refs.~\cite{RuizArriola:2016ihz,Troshin:2016frs,Albacete:2016pmp,Broniowski:2017aaf,Broniowski:2017rhz,Troshin:2017ucy,Dremin:2018orv,Campos:2018tsb,Dremin:2018urc,Broniowski:2018xbg,Petrov:2018wlv,Dremin:2019tgm} for more recent theoretical discussions on this apparently rather  fundamental-looking nature of proton-proton scattering at LHC and asymptotic  energies.

\begin{table*}
\centering
\caption{Summary of cross-section results of TOTEM at $\sqrt{s} = 2.76 $ and $13$ TeV in $pp$ collisions~\cite{Nemes:2017gut,Antchev:2018rec,Antchev:2017dia,Antchev:2018edk}.}
\label{t:2.76} 
\begin{tabular}{ccccc}
\hline
$\sqrt{s}$ \, (TeV) & $\sigma_{tot}$ (mb) & $\sigma_{el}$ (mb) & $\sigma_{in}$ (mb) & $\sigma_{el}/\sigma_{tot}$ (\%) \\ \hline
2.76 & 84.7 $\pm$ 3.3 & 21.8 $\pm$ 1.4 & 62.8 $\pm$ 2.9 & 25.7 $\pm$  1.1  \\ \hline
13.0 & 110.6 $\pm$ 3.4 & 31.0 $\pm$ 1.7 & 79.5 $\pm$ 1.8 & 28.1 $\pm$  0.9  \\ \hline
\end{tabular}
\label{t:sigma}
\end{table*}

\subsection{$B$ and $R$ measurements}
Recent TOTEM measurements of the nuclear slope parameter $B$
and the diffractive maximum-to-minimum ratio $R$ are 
summarized and discussed in this sub-section.

\begin{figure*}
\centering
\includegraphics[width=0.80\textwidth]{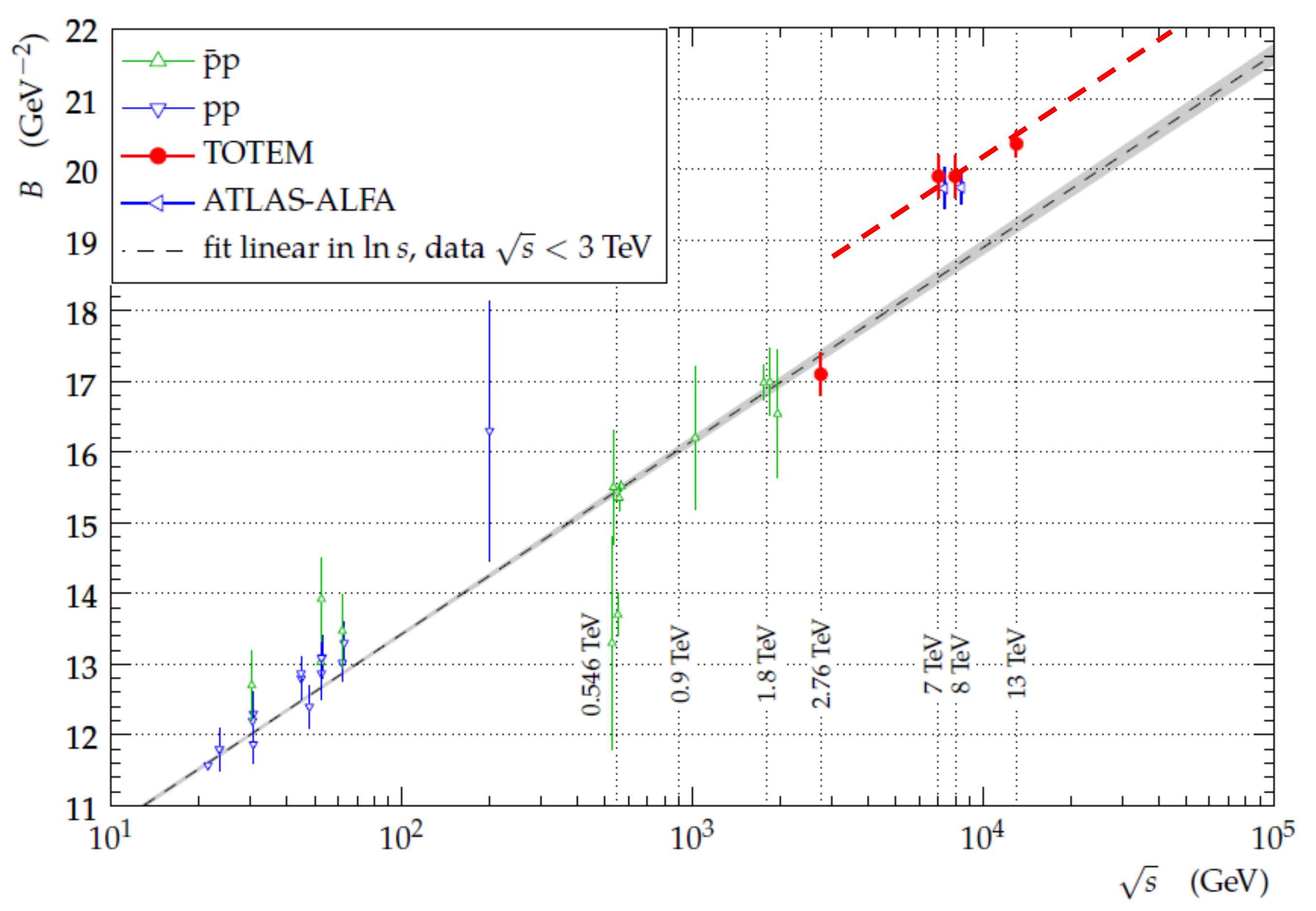}
\caption{Excitation function of the slope parameter $B$ in elastic proton-proton scattering. The TOTEM data at 2.76, 7, 8 and 13 TeV suggest the opening of a new channel 
between $\sqrt{s}$ = 2.76 and 7 TeV, as noted in  ref.~\cite{Antchev:2017dia} and indicated qualitatively by the red dashed line.}
\label{f:B(s)}       
\end{figure*}

The growth of $B$ as well as the growth of $\sigma_{\rm tot}$ with increasing collision energies $\sqrt{s}$ is characterizing the universal properties of proton-proton scattering and indicate the dominance of a colorless exchange. 
The most recent TOTEM measurements on the nuclear slope parameter at $\sqrt{s} = 2.76 $ and $13$ TeV
are summarized in Table~\ref{t:BR}.
It is quite remarkable that the excitation function of nuclear slope parameter $B(s)$ 
at $\sqrt{s} = 2.76$  TeV follows closely the trends of nuclear slopes measured before below the TeV energy scale, as shown on the summary plot for $B(s)$ on Figure~\ref{f:B(s)}.

Earlier TOTEM results of the nuclear slope parameter $B$  measured at 
$\sqrt{s} = 7 $ and $8$ TeV corresponded to results above the low-energy trend, while the $B$ value measured at 2.76 TeV follows the low-energy trends. The most recent TOTEM result for $B$ at $\sqrt{s} = 13$ TeV confirms the new trend seen already at 7 and 8 TeV.
Thus TOTEM  result on $B(s)$ suggests the opening of a new physics channel or a new domain
of proton-proton scattering, that starts slightly above 2.76 TeV but below 7 TeV. These results are fully consistent with the threshold behaviour of the elastic to total cross-section ratio,
that reaches the treshold called refractive scattering domain and the development of a hollow inside the proton at $\sqrt{s_{\rm th}} \approx 2.76 - 4.0$ TeV, as indicated on the right panel of Figure~\ref{f:stot-el-inel-s}.

\begin{table*}
\centering
\caption{Summary of $B$ and $R$ measurements of TOTEM at $\sqrt{s} = 2.76 $ and $13$ TeV in $pp$ collisions~\cite{Nemes:2017gut,Antchev:2018rec,Antchev:2017dia,Antchev:2018edk}.}
\label{t:2.76}       
\begin{tabular}{ccc}
$\sqrt{s}$\, (TeV) & $B$ ( GeV$^{-2}$ ) & R \\ \hline
2.76 & 17.1 $\pm $ 0.3 & 1.7  $\pm 0.2$ \\ \hline
13.0 & 20.40 $\pm  0.01^{\rm syst}  \pm {0.002}^{\rm stat}$  & 1.77  $\pm 0.01^{\rm stat}$ \\ \hline
\end{tabular}
\label{t:BR}
\end{table*}


\subsection{Differential elastic cross section measurements at $\sqrt{s}= 2.76$ and $13$ TeV}

Detailed measurements of the differential cross-section of elastic $pp$ and $p\overline{p}$ measurements
indicate that the nearly exponential cone behaviour is first of all only approximately exponential,
precision measurements reveal a non-exponential component.
Such a non-exponential feature of the elastic differential cross-section was reported first, as far as we know,
in high statistics $\pi^{+}p$, $\pi^{-}p$ and $pp$ collisions at an incident-beam momentum of $200$ GeV/c in the FNAL - E - 0069 experiment~\cite{Schiz:1979rh} and was also  reviewed in ref.
~\cite{Goulianos:1982vk}.
TOTEM found a significant, more than 7$\sigma$ effect, in high precision measurements
of the elastic $pp$ scattering at $\sqrt{s}= 8$ TeV~\cite{Antchev:2015zza,Antchev:2016vpy}.
The analysis of the hadronic part of the scattering amplitude outside the CNI region 
resulted in an observation of the non-exponential diffractive cone effect also in $pp$ elastic scattering at the currently highest available energy of $\sqrt{s} = 13$ TeV, see Table 5 and Figure 13 of ref. ~\cite{Antchev:2017yns}.

This non-exponential feature is followed by a diffractive minumum and a diffractive maximum in elastic
$pp$ collisions. It is important to note, that no secondary minimum or maximum structure is observed, although the TOTEM acceptance extends to several times $t_{\rm min}$, the $t$-value corresponding to the diffractive minimum both at 7 and 13 TeV. According to the investigations of Czyz and Maximon
of elastic scattering of composite particles in multiple diffraction theory, 
a single diffractive minimum corresponds to $(2,2)$ elastic scattering: if the symmetric scattering objects contain more than two sub-structures, more than a single diffractive minimum develops
~\cite{Czyz:1969jg}. These ideas were elaborated for asymmetric internal structures in the framework
of the quark-diquark model of protons by Bialas and Bzdak~\cite{Bialas:2006qf,Nemes:2015iia}. This model came in two variants, in one case the proton is assumed to be a weakly  bound state of a quark or diquark, abbreviated as $p = (q,d)$, but the internal structures of the quarks and diquarks are unresolved. In this case the $pp$ elastic scattering develops indeed a single minimum, in agreement
with the observations from the ISR energy of $\sqrt{s}= 23.5$ GeV up to the LHC energy of
7 TeV, after a small real part is added to the elastic scattering amplitude in a unitarized way
~\cite{Nemes:2015iia}. If the diquark were resolved as a weekly bound state of two quarks, $d = (q,q)$
and $p=(q,(q,q))$ t at least two diffractive minima would become observable in the $0 \le -t \le 2.5$ GeV$^2$ kinematic domain, while experimentally only a single diffractive minimum is observed. This dip region is followed by a diffractive maximum or bump, continued in  a monotonically decreasing and apparently structureless tail. 

\begin{figure*}
\centering
\includegraphics[width=0.51\textwidth]{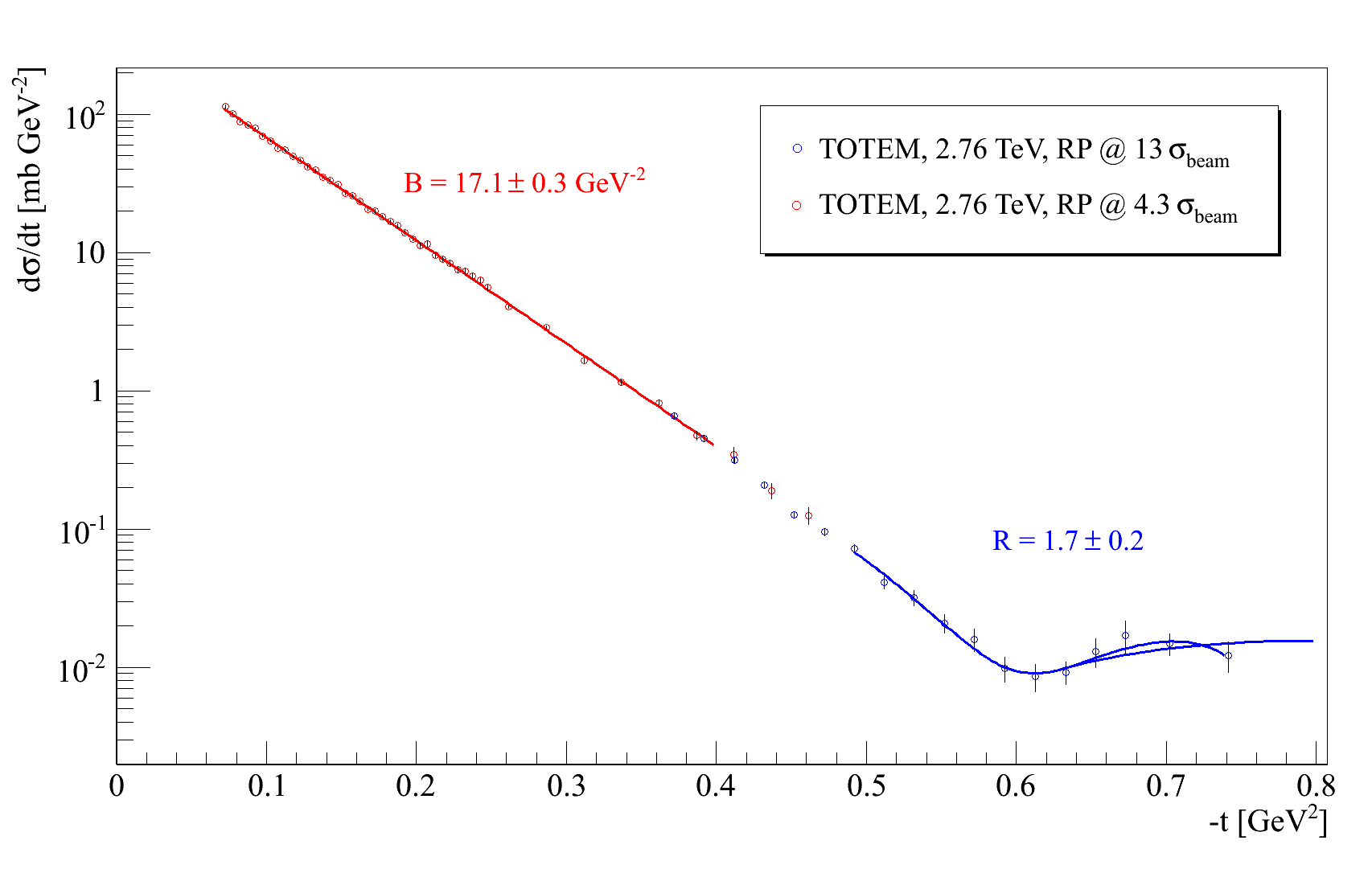}
\raisebox{0.75ex}{\includegraphics[width=0.45\textwidth]{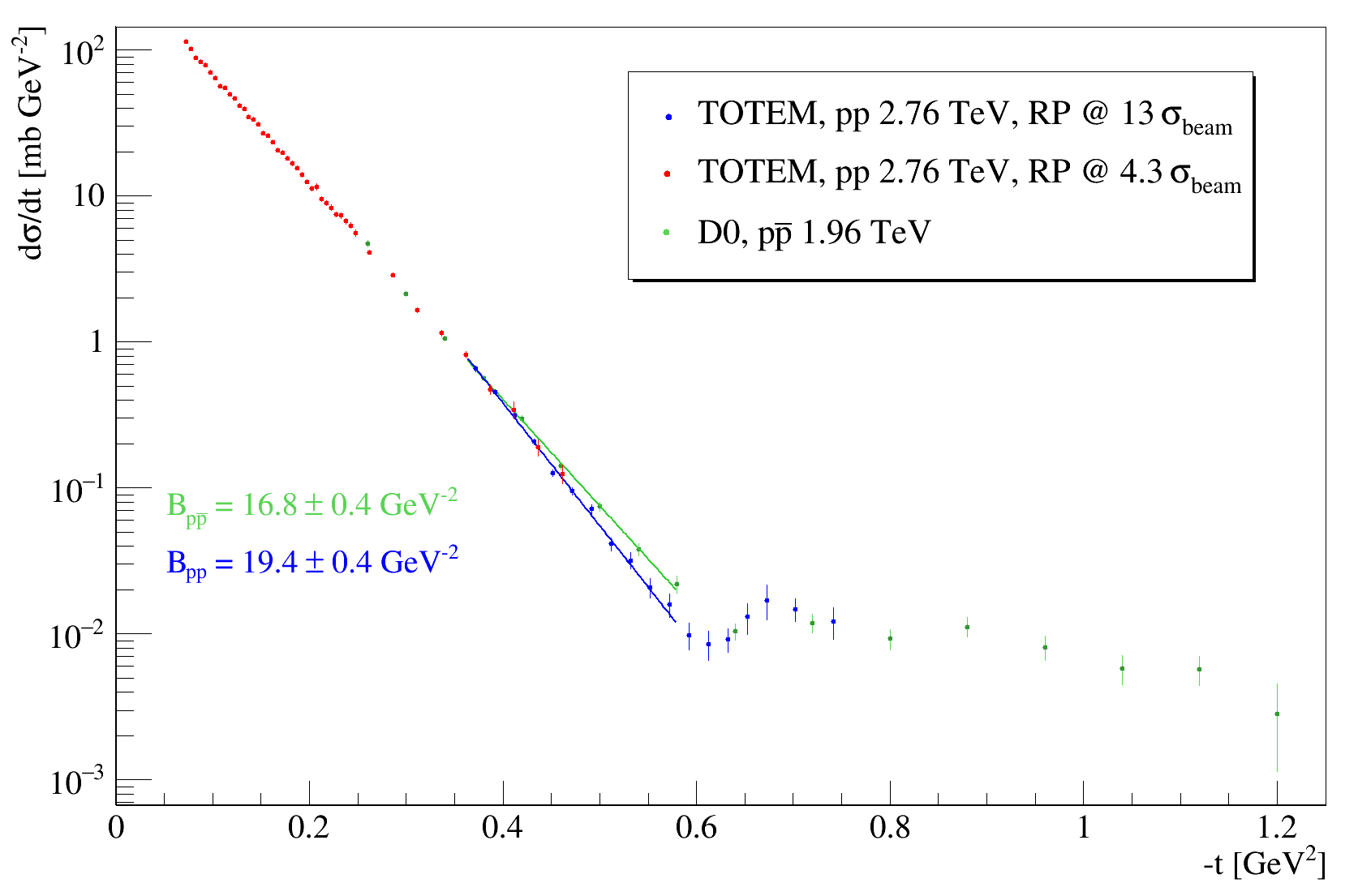}}
\caption{
{\bf Left panel:} 
Differential cross-section of elastic $pp$ scattering measured at the  LHC energy of $\sqrt{s} = 2.76$ TeV by TOTEM, as compared to third order polynomial fits in the dip and bump region~\cite{Antchev:2018rec}. The nuclear slope parameter is $B = 17.1 \pm 0.3$ GeV$^{-2}$, the maximum/minimum ratio is $R = 1.7 \pm 0.2$ for this $pp$ dataset.
{\bf Right panel:}
Differential cross-section of elastic $pp$ scattering measured at the  LHC energy of $\sqrt{s} = 2.76$ TeV by TOTEM~\cite{Antchev:2018rec}, as compared to D0 data on elastic proton-antiproton scattering at the Tevatron energy of
$\sqrt{s} = 1.96$ TeV~\cite{Abazov:2012qb}. The nuclear slope parameter "swings" ie increases at about $-t \approx 0.4$ GeV$^{-2}$ up to  $B_{pp} = 19.4 \pm 0.4$ GeV$^{-2}$ in $pp$ at $\sqrt{s} = 2.76 $ TeV, while it remains within errors constant in the $p\overline{p}$
dataset of D0 at the comparable $\sqrt{s} = 1.96 $ TeV in the same $-t$ region, $B_{p\overline{p}} = 16.8 \pm 0.4$ GeV$^{-2}$. The maximum/minimum ratio is $R = 1.0 \pm 0.0$ for the D0 $p\overline{p}$ dataset.
Figure from ref.~\cite{Antchev:2018rec}.
}
\label{f:dsdt-2.76} 
\end{figure*}
\begin{figure*}
\centering
\includegraphics[width=0.48\textwidth]{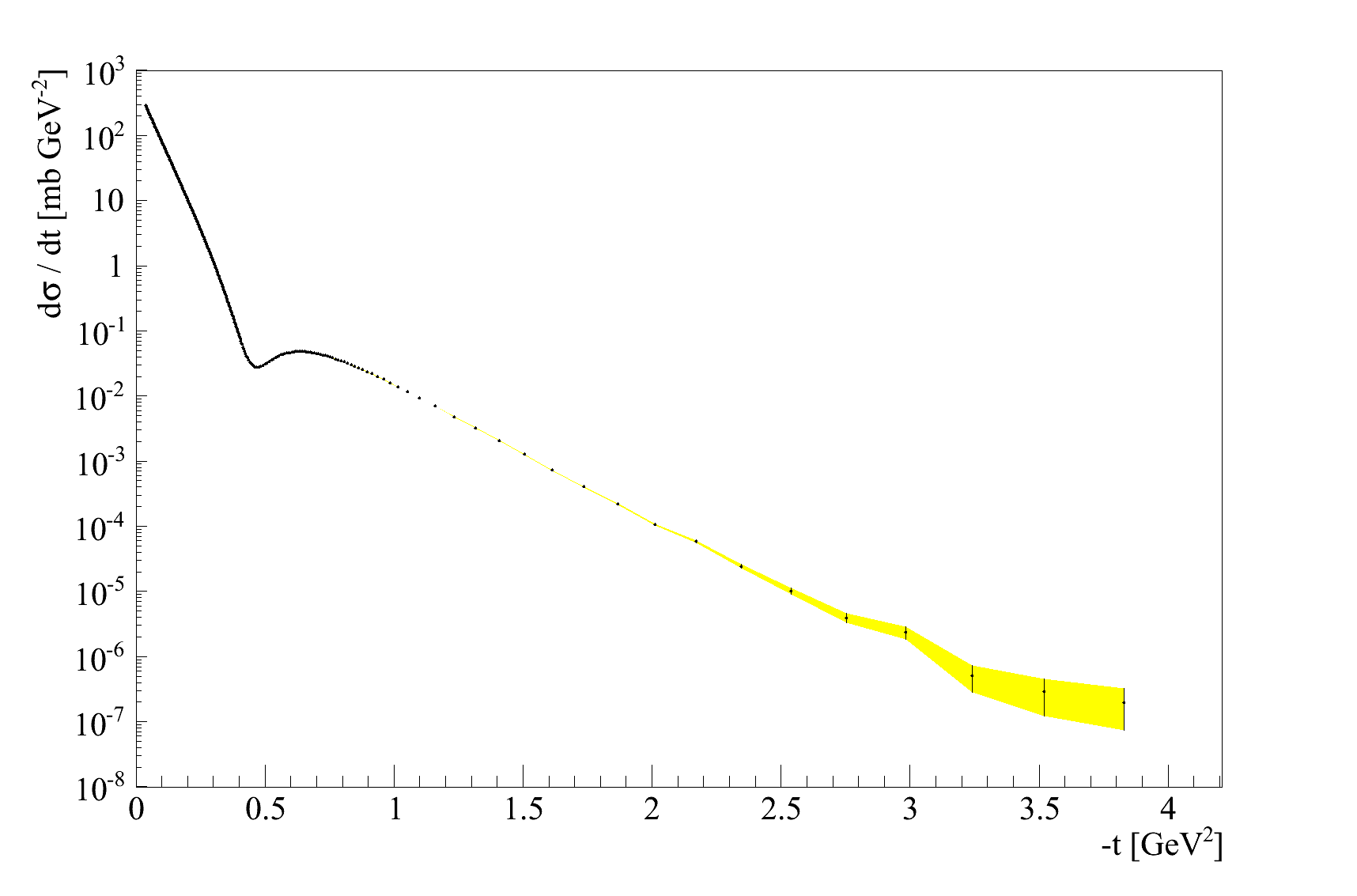}
\includegraphics[width=0.50\textwidth]{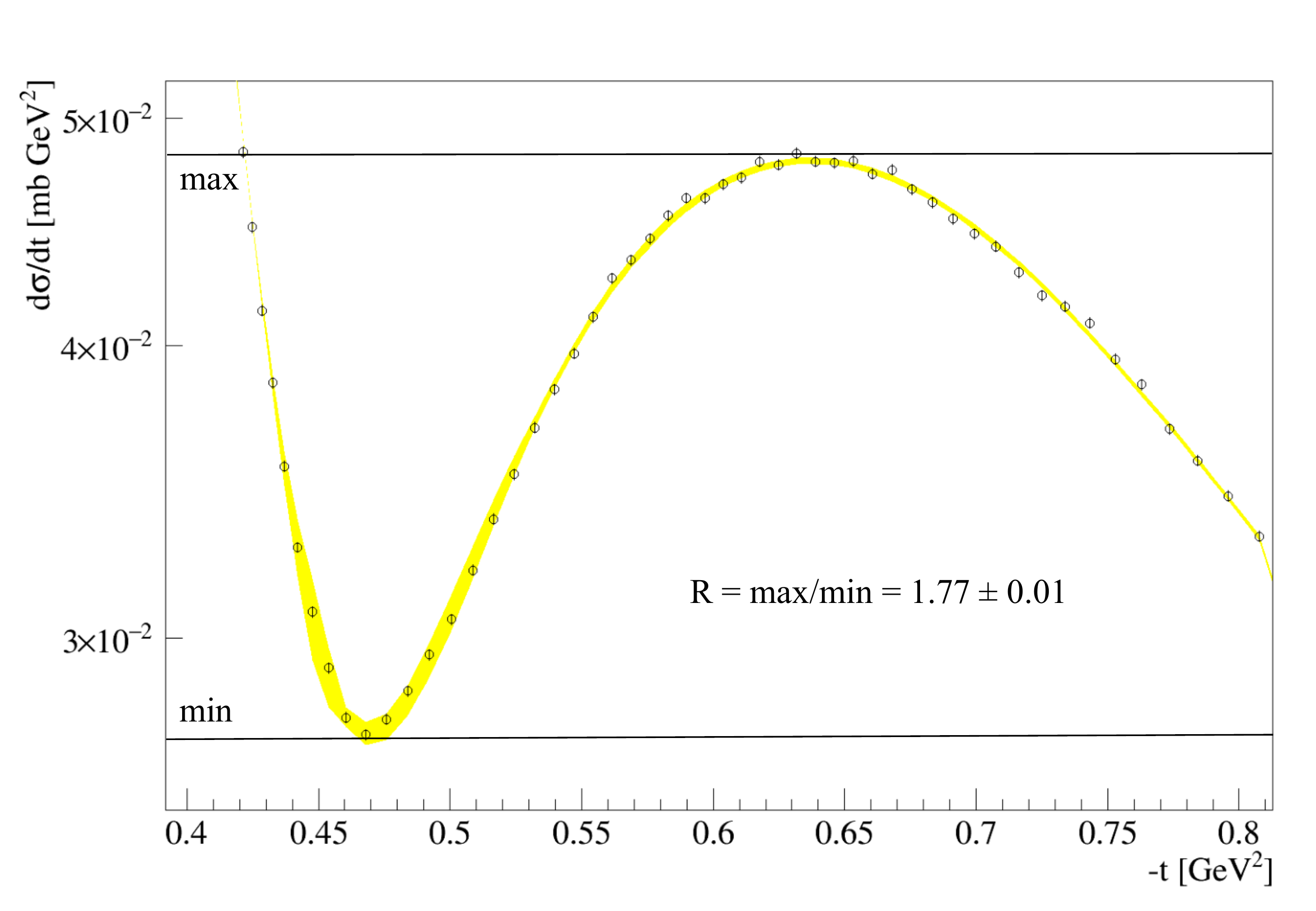}
\caption{
{\bf Left panel:}
TOTEM data for the differential cross-section of elastic $pp$ scattering at $\sqrt{s} = 13$ TeV, from ref. 
~\cite{Antchev:2018edk}. The statistical and $|t|$-dependent correlated systematic uncertainty envelope is indicated by a yellow band.
{\bf Right panel:}
The same TOTEM data for the differential cross-section of elastic $pp$
scattering at $\sqrt{s} = 13$ TeV as on the left panel,
but zooming in to the dip-and-bump region.
}
\label{f:dip-bump-13-TeV}       
\end{figure*}

In the region of the diffractive minimum and maximum, a third order polynomial fit was utilized to extract
the value of the differential cross-section at the diffractive maximum and minimum.
The left panel of Fig.~\ref{f:dsdt-2.76} indicates
that the  ratio R of diffractive maximum to diffractive minimum can be determined in $pp$ reactions at $\sqrt{s} = 2.76 $ TeV
reasonably well.
This left panel of Fig.~\ref{f:dsdt-2.76} shows two different third order polynomial fits, that indicate that the ratio $R$ is rather stable for the choice of the fitting function, however the position of the diffractive maximum is rather uncertain and more data with better statistics, and if possible at larger values of $t$ are desirable to determine precisely the position of the diffractive maximum as a function of $t$. 

The right panel of Fig.~\ref{f:dsdt-2.76} also compares the differential cross-section measurement of TOTEM at $\sqrt{s} = 2.76$ TeV for elastic $pp$ scattering with the similar measurement of the D0 collaboration for $p\overline{p}$ elastic scattering at the slightly lower energy of  $\sqrt{s} = 1.96$ TeV~\cite{Abazov:2012qb}. This Figure indicates that $R =  1.0 \pm 0.1$ for  $p\overline{p}$ elastic scattering at Tevatron energies. The same plot also shows that the $t$-dependent slope parameter $B(t)$ is clearly different for proton-proton and proton-antiproton elastic scattering in the $-t\approx 0.4 $ GeV$^2$ region, the difference being 2.6 $\pm$ 0.56 GeV$^{-2}$,
a more than 4$\sigma$ effect.

Fig.~\ref{f:dip-bump-13-TeV} indicates TOTEM data for the differential cross-section of elastic $pp$ scattering at $\sqrt{s} = 13$ TeV, from ref. ~\cite{Antchev:2018edk}. 
The systematic error range is indicated with a yellow band. The bump/dip ratio is found to be
$ R  = 1.77 \pm 0.01$ which is significantly different from a value of approximately 1.0 $\pm$ 0.1, as seen for $p\bar{p}$ elastic scattering at  $\sqrt{s} = 1.96$ TeV. The deviation of $R$ in elastic $pp$ collisions from that of elastic $p\overline{p}$ collisions can be interpreted as an  Odderon effect, if one can verify  that the variation  of its excitation function $R(s)$ due to the change of the energy of the collisions can be shown to be negligibly small between 2.76 TeV, the lowest $\sqrt{s}$ value investigated by TOTEM at the LHC and 1.96 TeV, the highest value where elastic $p\overline{p}$ reactions have been measured by D0 at the Tevatron energies~\cite{Abazov:2012qb}. 

Recently, the sensitivity of  the $t$-dependent elastic slope parameter $B(t)$ to Odderon effects was pointed out in refs.~\cite{Csorgo:2018uyp,Csorgo:2018ruk,Csorgo:2019rsr}, 
based on a model-independent L\'evy expansion method. These observations were confirmed in a model-dependent calculation that uses the maximal Odderon picture~\cite{Martynov:2018sga}. 
Similar conclusions were obtained using the Reggeized versions of the  Phillips-Barger model,
in refs.~\cite{Ster:2015esa,Goncalves:2018nsp}. Both of these results emphasized the effects of the Odderon contribution, by filling up  the region of the diffractive minimum in
$p\overline{p}$ reactions, when the calculations are performed in the LHC energy range.

Let us close the discussion of the differential cross-section of elastic $pp$ scattering at the LHC energy range by pointing out that Brodsky and Farrar predicted the asymtotic, large $s$ dependence of the differential cross-section at a fixed value of $t$ in ref.~\cite{Brodsky:1973kr}.
Such a behaviour can be readily converted to the large-$t$
asymptotic behaviour of the differential cross-section at a fixed value of $s$ as $\frac{d\sigma}{dt} \propto t^{-n}$, where the exponent $n = NDF-2$ corresponds to the internal degrees of freedom
in the incoming and outgoing particles. If the proton is a bound state of $3$ dressed quarks,
$p = (q,q,q)$ then the number of degrees of freedom is $NDF = 4\times 3 = 12$ and the exponent
$n$ is expected to be of the order of $10$. Currently at the largest available $t$ range at 
$\sqrt{s} = 13$ TeV, the exponent still seems to be fit range dependent, with approximate values
of the order of 10.

\section{Summary and conclusions}
\label{s:summary}
This manuscript reviewed the most recent results of the CERN LHC experiment TOTEM, achieved at the center-of-mass energy scales of $\sqrt{s} = 2.76$ and $13$ TeV. 

A clear experimental observation of a threshold effect is reported on the collision energy dependence of the nuclear slope parameter $B(s)$, that is found to undergo an abrupt increase in the energy range
between $\sqrt{s} = 2.76$ TeV and $7$ TeV. A similar but more indirect threshold effect is also
reported in the energy dependence of the ratio of the elastic to the total cross-section, $\sigma_{\rm el}(s)/\sigma_{\rm tot}(s)$, which seems to pass the important threshold of 1/4 also in the same energy region. Several theoretical considerations suggest that passing this threshold may result in a fundamental change in the shadow profile of proton-proton collisions, corresponding to the probability distribution of inelastic collisions in the impact parameter space.

Odderon effects were first identified by TOTEM in the $\sqrt{s}$ dependent
($\sigma_{\rm tot}$, $\rho$) excitation functions.
Theoretical models including the effects of the Odderon~\cite{Lukaszuk:1973nt,Avila:2006wy,Martynov:2017zjz}, have predicted the observed effects and were able to describe both the $pp$ TOTEM data and the D0 data of $p\overline{p}$ on the TeV scale.
As far as we know, there are no models which are able to describe these data without the effects of the Odderon exchange~\cite{Cudell:2002xe,Khoze:2017swe,Levin:1990gg}.

Subsequently,  even more significant Odderon effects were identified by TOTEM
in the shape analysis of the differential cross-section.
At each of the LHC energies of 13,  7 TeV and 2.76 TeV, the diffractive minimum and maximum has been observed  by TOTEM, with a fairly energy independent maximum to minimum ratio of $R = 1.77 \pm 0.01$,
$1.7 \pm 0.1$ and $1.7 \pm 0.2$, respectively. Thus the diffractive minimum and maximum ratio 
is apparently a permanent structure with approximately constant magnitude in $pp$ 
elastic scattering at LHC energies, and such a structure is apparently missing in $p\overline{p}$ collisions at the TeV scale~\cite{Abazov:2012qb}. Therefore, unless something unknown happens between $\sqrt{s} = $2.76 TeV and 1.96 TeV, the 
difference between the shape of the $pp $ and $p\overline{p}$ differential cross-sections
provides a promising signal of a  crossing-odd component in the forward scattering amplitude, corresponding to a predominantly 3-gluon bound state exchange in the t-channel of the proton-proton elastic scattering~\cite{Antchev:2018rec}. 
The clarification of the significance of these effects is a subject of a D0-TOTEM common publication, which is in preparation at the time of the closing of this manuscript.

\section*{Acknowledgments:}
T. Cs. would like to express his gratitude to the Organizers of ISMD 2018 for an invitation, partial support and in particular for  an outstanding, inspiring and useful meeting and to thank G. Gustafson and R. Pasechnik for inspiring discussions and kind hospitality at the University of Lund, Sweden. T. Cs. was partially supported by the COST Action CA15213, THOR Project of the European Union, and by the Hungarian NKIFH grants FK-123842 and FK-123959.
This research has been partially  supported by the Institutions of the  TOTEM Collaboration, and the US NSF, the Magnus Ehrnrooth Foundation (Finland), the Waldemar von Frenckell Foundation (Finland), the
Academy of Finland, the Finnish Academy of Science and Letters, the Circles of Knowledge Club (Hungary) and the OTKA NK 101438 and the EFOP-3.6.1-16-2016-00001 grants (Hungary).

\end{document}